\documentclass[10pt,conference]{IEEEtran}

\ifCLASSINFOpdf
   \usepackage[pdftex]{graphicx}
\else
   \usepackage[dvips]{graphicx}
\fi

\ifCLASSOPTIONcomsoc
  \usepackage[caption=false,font=normalsize,labelfont=sf,textfont=sf]{subfig}
\else
  \usepackage[caption=false,font=footnotesize]{subfig}
\fi

\usepackage{footnote}
\usepackage{nopageno}
\usepackage{float}
\usepackage{url}
\usepackage{amssymb,amsfonts}
\usepackage{amsmath}

\usepackage{graphicx}
\usepackage{mathptmx}  
\usepackage{epstopdf}
\usepackage{textcomp}
\usepackage{xcolor}
\usepackage{xspace}
\usepackage[absolute,showboxes]{textpos}

\usepackage{multirow}
\usepackage{multicol}
\usepackage{array} 
\usepackage{cite}
\usepackage{algorithm,algorithmic}
\usepackage{amsmath}
\usepackage[cmintegrals]{newtxmath}
\usepackage{balance}  

\usepackage[utf8]{inputenc}
\usepackage{blindtext}
\usepackage{csquotes}

\begin{document}
\pagenumbering{gobble}

\title{DeepFaith: Evidence-Grounded LLMs for Faithful Incident Reporting in Multi-Stage APT Defense}

\author{

Trung V. Phan\IEEEauthorrefmark{2}, Tri Gia Nguyen\IEEEauthorrefmark{4} and Thomas Bauschert\IEEEauthorrefmark{2} \\
\IEEEauthorblockA{\IEEEauthorrefmark{2}Chair of Communication Networks, Technische Universit{\"a}t Chemnitz,  09126 Chemnitz, Germany}
\IEEEauthorblockA{\IEEEauthorrefmark{4}Department of Information Assurance, FPT University, Da Nang 50509, Vietnam}
Email: trung.phan-van@etit.tu-chemnitz.de, tri@ieee.org, thomas.bauschert@etit.tu-chemnitz.de

%\\
%\textcolor{blue}{This paper is currently under review for IEEE CNSM 2026.}
}

\maketitle

\begin{abstract}
Advanced Persistent Threats (APTs) are difficult to detect and interpret due to their multi-stage and stealthy nature. While recent autonomous defense systems leverage provenance graphs and learning-based models for detection and mitigation, their outputs remain largely machine-oriented and difficult for analysts to interpret. Large language models (LLMs) offer a promising interface for report generation, but often produce hallucinated or weakly grounded content. In this paper, we propose DeepFaith, an evidence-grounded framework for faithful incident reporting in multi-stage APT defense. DeepFaith transforms structured outputs from autonomous defense and explainability modules into natural-language reports that are explicitly aligned with underlying system evidence. The framework integrates a unified evidence representation, evidence-grounded prompting, faithfulness-aware generation, and post-generation verification to ensure that all generated statements are supported. Experiments in a realistic enterprise testbed demonstrate that DeepFaith improves faithfulness from 0.68 to 0.92, reduces unsupported claims from 0.32 to 0.08, and increases temporal consistency from 0.6 to 0.88, while maintaining concise reports and lower error rates than existing template-based and LLM-based solutions. These results show that evidence-grounded generation enables reliable, interpretable, and actionable reporting for security operations centers.
\end{abstract}

\IEEEpeerreviewmaketitle

\pagestyle{headings}
\setcounter{page}{1}
\pagenumbering{arabic}

\section{Introduction}
Advanced Persistent Threats (APTs) \cite{rehman2024flash} represent one of the most challenging classes of cyber attacks due to their stealthy, multi-stage, and long-term nature. Unlike conventional attacks, APTs unfold over extended periods, traverse multiple hosts, and progress through stages such as reconnaissance, initial compromise, lateral movement, and data exfiltration. Detecting and mitigating such attacks requires not only analyzing large volumes of heterogeneous telemetry data, but also reasoning over complex causal dependencies and temporal evolution across the system \cite{zhang2024survey}. Recent research has therefore shifted toward provenance-based and learning-driven approaches that model system behavior as structured graphs and leverage deep learning for detection \cite{StageFinder}. 

Building on these advances, autonomous cyber defense frameworks have been proposed to perform adaptive mitigation in real time \cite{DeepAir}. In particular, our prior stage-aware system \textit{DeepStage}~\cite{deepstage} combines graph neural networks (GNNs) with reinforcement learning (RL) to infer attack progression and select defense actions. In addition, our prior complementary work \textit{DeepXplain}~\cite{deepxplain} introduces explainability mechanisms that extract structural, temporal, and decision-level evidence from these models. While these approaches improve detection accuracy and policy effectiveness, their outputs remain largely machine-centric, consisting of latent embeddings, probability distributions, and explanation signals. As a result, human analysts in security operations centers (SOCs) must still manually interpret and synthesize this information, slowing response.

At the same time, large language models (LLMs) have emerged as a promising tool for assisting SOC workflows, including alert triage, log analysis, and incident reporting~\cite{habibzadeh2025socsurvey,jaffal2025llmcyber}. However, recent studies show that unconstrained LLMs often produce hallucinated or unverifiable content, especially in complex and safety-critical domains~\cite{kramer2025incident}. In the context of cyber defense, such errors can lead to incorrect conclusions and inappropriate mitigation actions. This limitation suggests that LLMs should not be used as standalone reasoning engines, but rather as communication layers that translate verified evidence into human-readable outputs.

These observations reveal a fundamental gap: while upstream systems can detect, explain, and respond to APT attacks, there is no principled mechanism to convert their outputs into concise, faithful, and analyst-friendly reports. Existing reporting approaches are either template-based, which lack flexibility, or LLM-based, which lack grounding and reliability \cite{yang2025llmknowledge,structuredllm2026survey}. Bridging this gap is critical for enabling effective human–AI collaboration in SOC environments.

To address this challenge, we propose DeepFaith, a framework for evidence-grounded incident report generation. DeepFaith takes structured outputs from autonomous defense and explainability modules and transforms them into natural-language reports that are both informative and verifiably supported by evidence. The key idea is to treat report generation as a constrained conditional generation problem, where the language model is guided by structured evidence and explicitly evaluated for faithfulness. 

DeepFaith introduces three main contributions. \textit{First}, we define a unified evidence representation that integrates stage estimation, provenance structure, temporal information, and action rationale into a common interface. \textit{Second}, we propose a faithfulness-aware generation framework incorporating evidence-grounded prompting, faithfulness regularization, temporal consistency constraints, and confidence-aware expression. \textit{Third}, we design a verification and regeneration mechanism that evaluates reports against underlying evidence and filters unsupported outputs. We evaluate DeepFaith in a realistic enterprise testbed building on our prior \textit{DeepStage} \cite{deepstage} and \textit{DeepXplain} \cite{deepxplain} frameworks, using multi-stage APT scenarios generated via an adversarial emulation framework \cite{mitre_caldera}. Results show that DeepFaith significantly improves faithfulness, reduces hallucination, and enhances temporal alignment compared to template-based and LLM-based baselines.

\section{Related Work}

\subsection{Autonomous APT Detection and Defense}
Advanced Persistent Threats (APTs) are difficult to detect due to their long duration, multi-host propagation, and weak local indicators. As a result, research has moved beyond signature-based methods toward graph-based and learning-driven approaches that capture causal dependencies and temporal evolution. Surveys highlight that modern APT defense increasingly relies on integrated telemetry, attack-stage reasoning, and adaptive response~\cite{alshamrani2019survey,zhang2024survey}.

Provenance graphs have emerged as a powerful representation for modeling system behavior, preserving both structural and temporal relationships among entities. Systems such as FLASH demonstrate that graph representation learning over provenance data can significantly improve intrusion detection performance in enterprise environments~\cite{rehman2024flash}. However, practical deployment remains challenging due to scalability and attribution limitations~\cite{bilot2025simpler}.

Reinforcement learning (RL) has been widely explored for adaptive cyber defense, enabling sequential decision-making under uncertainty in dynamic environments~\cite{nguyen2021drlcyber,vyas2025realistic}. Recent work incorporates risk-aware or entity-centric state representations for effective mitigation planning~\cite{le2024automated,thompson2024entity}. More recently, our prior stage-aware framework \textit{DeepStage}~\cite{deepstage} integrates provenance graphs with attack-stage estimation to guide hierarchical RL policies. While these approaches improve defense effectiveness, their outputs remain largely machine-oriented, limiting direct usability for human analysts.

\subsection{Explainability for Reinforcement Learning and Graph-Based Security}
Interpretability in sequential decision-making has motivated increasing interest in explainable reinforcement learning (XRL). Existing approaches focus on feature attribution, policy summarization, and explanation generation, but are often loosely coupled with decision-making and remain domain-agnostic~\cite{milani2024xrl}. This limitation is critical in cyber defense, where high-impact mitigation actions require verifiable justification.

Explainability for graph neural networks (GNNs) has also been extensively studied. Methods such as GNNExplainer identify important subgraphs by learning masks over nodes and edges~\cite{ying2019gnnexplainer}, while recent surveys emphasize evaluation criteria such as correctness, robustness, and compactness~\cite{li2025gnnsurvey,dai2024trustworthy}. Although effective for static prediction tasks, these methods are less suited for dynamic, action-driven environments.

To address this, our prior \textit{DeepXplain}~\cite{deepxplain} integrates structural, temporal, and policy-level explanations into autonomous defense through evidence alignment and confidence-aware learning. While this improves interpretability at the model level, the outputs remain machine-centric and do not directly support analyst-facing communication. 

\subsection{LLMs for Security Operations and Incident Reporting}
Large language models (LLMs) are increasingly explored for SOC tasks such as alert triage, log analysis, and incident reporting. Recent surveys highlight their potential for workflow acceleration, while emphasizing risks related to hallucination, inconsistency, and unverifiable reasoning~\cite{habibzadeh2025socsurvey,jaffal2025llmcyber}. In particular, Kramer \textit{et al.} show that fully autonomous LLM-based incident summaries often omit key details and introduce factual errors, whereas human-in-the-loop approaches yield more reliable results~\cite{kramer2025incident}. This suggests that LLMs are better suited as communication layers rather than standalone reasoning systems.

\subsection{Grounded Generation from Structured Evidence}
A growing body of work studies text generation from structured inputs such as databases and knowledge graphs. These approaches emphasize not only fluency, but also faithfulness, schema adherence, and execution-grounded correctness~\cite{yang2025llmknowledge,structuredllm2026survey}. Such considerations are particularly important in SOC environments, where unsupported claims can lead to operational risk. However, existing grounded-generation methods are largely domain-agnostic and do not account for the specific characteristics of cyber defense outputs, such as stage distributions, provenance subgraphs, temporal evidence, and action rationales. Conversely, prior cyber-defense research focuses on detection and explanation, but not on grounded natural-language reporting.

\subsection{Research Gap}
In summary, prior work shows that (\textit{i}) provenance-based and stage-aware models, including our prior DeepStage framework, provide a strong foundation for APT defense~\cite{rehman2024flash,deepstage}, (\textit{ii}) explainability methods, including our prior DeepXplain framework, expose structural and temporal evidence but remain machine-centric~\cite{milani2024xrl,li2025gnnsurvey,deepxplain}, and (\textit{iii}) LLMs are effective for analyst-facing communication only when properly grounded~\cite{kramer2025incident,habibzadeh2025socsurvey}. Therefore, DeepFaith is designed at this intersection, enabling faithful and interpretable report generation from structured defense outputs.

\begin{figure*}
    \centering
    \includegraphics[width=1.0\linewidth]{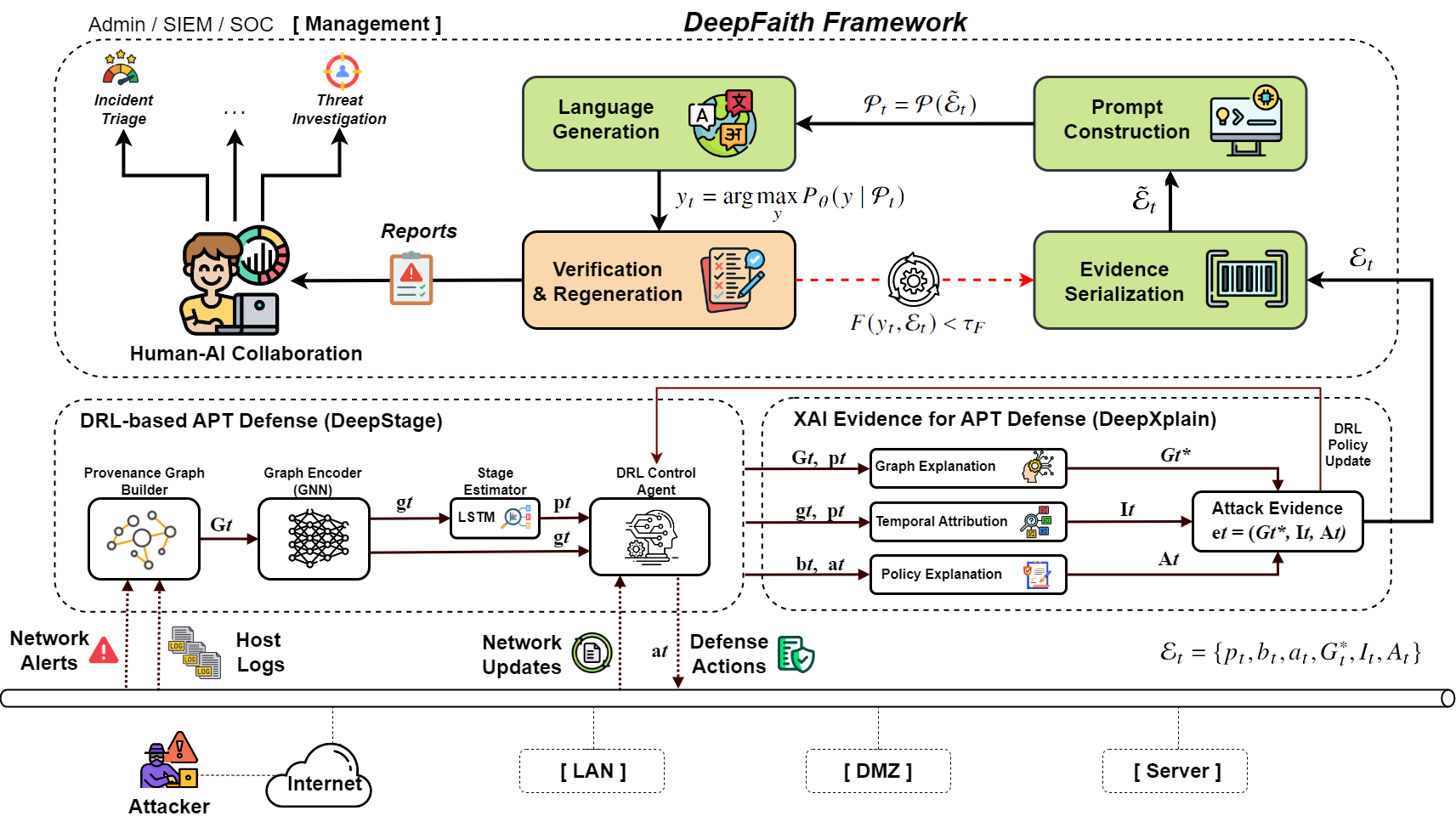}
    \caption{Data and control flow of the proposed DeepFaith framework building upon the DeepStage \cite{deepstage} and the DeepXplain \cite{deepxplain} solutions.}
    \label{fig:DeepFaith_framework}
\end{figure*}

\section{System Model}

\subsection{Autonomous Defense and Explanation Pipeline}
We build on our prior APT defense architecture \textit{DeepStage}~\cite{deepstage} together with our prior explainability pipeline \textit{DeepXplain}~\cite{deepxplain}. As shown in Fig.~\ref{fig:DeepFaith_framework}, enterprise telemetry from host and network monitoring is transformed into a dynamic provenance graph $G_t = (V_t, E_t)$ at time $t$, where nodes represent system entities (e.g., processes, files, sockets) and edges encode causal interactions such as execution, access, and communication. This representation preserves structural and temporal dependencies essential for modeling multi-stage APT behavior in the enterprise network environment.

A graph neural network encodes $G_t$ into an embedding $g_t = f_{\mathrm{GNN}}(G_t)$, summarizing system interactions in a compact latent vector. Temporal dynamics are captured by a recurrent stage estimator,
$
p_t = f_{\mathrm{LSTM}}(g_1, \dots, g_t),
$
where $p_t \in \mathbb{R}^K$ denotes the probability distribution over $K$ attack stages. To handle partial observability, a belief state $b_t$ is maintained as
$
b_t = f_{\mathrm{LSTM}}(b_{t-1}, o_t), \quad
o_t = [g_t, p_t, a_{t-1}],
$
which integrates current observations and past actions. A policy then selects a mitigation action $a_t \sim \pi_\theta(a \mid b_t)$ for adaptive defense~\cite{deepstage}.

To improve interpretability, an explainability module~\cite{deepxplain} operates on the system state $\mathcal{X}_t = \{G_t, g_t, p_t, b_t, a_t\}$ and produces an explanation signal
$
e_t = (G_t^*, I_t, A_t),
$
where $G_t^*$ is a salient subgraph capturing critical interactions, $I_t$ identifies influential time steps, and $A_t$ provides the rationale for action selection. These components expose structural, temporal, and decision-level evidence for downstream use.

\subsection{Evidence Representation}

The outputs of the defense and XAI modules are heterogeneous and not directly suitable for reporting. To bridge this gap, we define a unified evidence representation:
\begin{equation}
\mathcal{E}_t = \{p_t, b_t, a_t, G_t^*, I_t, A_t\}.
\end{equation}

This representation aggregates the key information required to describe system state and defense decisions: $p_t$ encodes stage probabilities, $b_t$ captures latent context, $a_t$ specifies the selected action, $G_t^*$ provides structural evidence, $I_t$ captures temporal progression, and $A_t$ explains the action rationale. As such, $\mathcal{E}_t$ serves as a structured interface between the autonomous defense system and the DeepFaith framework.

\section{Problem Formulation}

Given structured evidence $\mathcal{E}_t$, the goal is to generate a natural-language incident report $y_t$ that accurately reflects the system state and defense decision:
\begin{equation}
y_t \sim P(y \mid \mathcal{E}_t).
\end{equation}
The report $y_t$ consists of three components: (\textit{i}) an \textit{incident summary} describing the attack stage and risk level, (\textit{ii}) an \textit{attack timeline} capturing key events, and (\textit{iii}) a \textit{mitigation rationale} explaining the selected action. Unlike standard text generation, outputs must remain grounded in $\mathcal{E}_t$, ensuring that all statements are supported by model-derived evidence.

\subsection{Faithfulness Constraint}

To enforce grounding, we define a faithfulness function measuring alignment between $y_t$ and $\mathcal{E}_t$. Let $\phi(y_t)$ extract atomic claims from the report, and let $\psi(\mathcal{E}_t)$ denote evidence-supported facts derived from: stage information from $p_t$, events from $G_t^*$, timestamps from $I_t$, action rationale from $A_t$. A claim $c \in \phi(y_t)$ is supported if it matches a fact in $\psi(\mathcal{E}_t)$. The faithfulness score is:
\begin{equation}\label{Faithfulnes_score}
F(y_t, \mathcal{E}_t) =
\frac{1}{|\phi(y_t)|}
\sum_{c \in \phi(y_t)} \mathbf{1}[c \in \psi(\mathcal{E}_t)].
\end{equation}
This metric captures the proportion of evidence-supported claims, penalizing unsupported content.

\subsection{Temporal Consistency}
The report must preserve the temporal structure of the attack. Let $\pi = (e_1,\dots,e_n)$ denote the generated event sequence. Temporal consistency is defined with respect to key timestamps $I_t$:
\begin{equation}\label{Temporal_consistency}
C_{\mathrm{time}} =
\frac{1}{n}
\sum_{i=1}^{n} \mathbf{1}[\mathrm{timestamp}(e_i) \in I_t].
\end{equation}
Chronological ordering must satisfy $\mathrm{timestamp}(e_i) \leq \mathrm{timestamp}(e_{i+1})$, and stage progression should not regress over time. These constraints ensure coherent and evidence-aligned timelines.

\subsection{Confidence-Aware Expression}
Let $c_t = \max_k p_t^{(k)}$ denote the stage confidence. This value regulates linguistic expression: high confidence permits assertive statements, while lower confidence encourages cautious phrasing. A mapping $u_t = f_{\mathrm{conf}}(c_t)$ is incorporated into prompting to align language with model uncertainty.

\subsection{Optimization Objective}

The objective combines generation quality with grounding constraints:
\begin{equation}\label{OptimizationProblem}
\mathcal{L} =
\mathcal{L}_{\mathrm{gen}} +
\lambda_1 (1 - F(y_t, \mathcal{E}_t)) +
\lambda_2 \mathcal{L}_{\mathrm{compact}},
\end{equation}
where $\lambda_1$ and $\lambda_2$ balance fluency, faithfulness, and conciseness.

The generation loss is:
\begin{equation}
\mathcal{L}_{\mathrm{gen}} =
- \sum_{i=1}^{|y_t|}
\log P_\theta(y_i \mid y_{<i}, \mathcal{P}_t),
\end{equation}
and the compactness term penalizes excessive length:
\begin{equation}\label{L_compact}
\mathcal{L}_{\mathrm{compact}} = \max(0, |y_t| - L_{\max}).
\end{equation}

This formulation encourages reports that are accurate, interpretable, and concise.

\section{Design of DeepFaith Framework}
In this section, we introduce the DeepFaith framework designed to transform structured evidence $\mathcal{E}_t$ into a human-readable incident report $y_t$ that is informative and faithful to the system state. As illustrated in Fig.~\ref{fig:DeepFaith_framework}, the framework acts as a module bridging machine-level outputs and analyst-facing reports. The transformation from $\mathcal{E}_t$ to $y_t$ is non-trivial because the input combines graph structures ($G_t^*$), probability distributions ($p_t$), temporal indices ($I_t$), and action explanations ($A_t$). To address this heterogeneity, DeepFaith decomposes report generation into four phases:
\begin{itemize}
    \item \textbf{Evidence serialization}: which converts structured evidence into normalized textual fields;
    \item \textbf{Evidence-grounded prompt construction}: which organizes serialized evidence into a constrained prompt;
    \item \textbf{Faithfulness-aware language generation}: which produces an initial report while enforcing grounding and consistency constraints;
    \item \textbf{Verification and regeneration}: which evaluates the generated report against the structured evidence and regenerates outputs that do not satisfy faithfulness requirements.
\end{itemize}

\subsection{Evidence Serialization}
The first phase converts the structured evidence $\mathcal{E}_t = \{p_t, b_t, a_t, G_t^*, I_t, A_t\}$ into a normalized intermediate representation $\tilde{\mathcal{E}}_t = \mathcal{S}(\mathcal{E}_t)$, where $\mathcal{S}(\cdot)$ is a deterministic serialization function. The purpose of serialization is to transform heterogeneous inputs into semantically meaningful textual fields that can be consumed by a language model. Formally, the serialized representation can be expressed as:
\begin{equation}
\tilde{\mathcal{E}}_t = \{s_t, \tau_t, \gamma_t, \alpha_t\},
\end{equation}
where $s_t$, $\tau_t$, $\gamma_t$, and $\alpha_t$ denote stage, temporal, structural, and action-level evidence fields, respectively.

This step is essential because it creates a stable and interpretable interface between the autonomous defense model and the language model. It also ensures that prompting is based on normalized evidence rather than raw heterogeneous outputs.

\subsection{Evidence-Grounded Prompt Construction}
The serialized evidence $\tilde{\mathcal{E}}_t$ is then converted into a structured prompt $\mathcal{P}_t = \mathcal{P}(\tilde{\mathcal{E}}_t)$, which serves as the conditioning context for the language model.
The prompt contains three main components:
\begin{itemize}
    \item \textit{Task instruction}: which specifies that the model must generate an incident report;
    \item \textit{Evidence fields}: which provide the serialized stage, temporal, structural, and action information;
    \item \textit{Grounding constraints}: which explicitly instruct the model to avoid unsupported claims and remain faithful to the provided evidence.
\end{itemize}
Formally, the prompt can be represented as:
\begin{equation}
\mathcal{P}_t = [\mathcal{I}, s_t, \tau_t, \gamma_t, \alpha_t],
\end{equation}
where $\mathcal{I}$ denotes the instruction template.

This prompt design imposes a strong inductive bias on the language model: rather than generating free-form text from loosely related context, the model is explicitly guided to produce outputs grounded in the evidence fields.

\subsection{Faithfulness-Aware Language Generation}
Given the structured prompt $\mathcal{P}_t$, DeepFaith generates the incident report through a conditional language model:
\begin{equation}
y_t \sim P_\theta(y \mid \mathcal{P}_t),
\end{equation}
where $\theta$ denotes the model parameters. To align language generation with the optimization objective defined in Equation \eqref{OptimizationProblem}, DeepFaith incorporates three complementary mechanisms.

\subsubsection{Faithfulness Regularization}
The faithfulness loss is defined as:
\begin{equation}
\mathcal{L}_{\mathrm{faith}} = 1 - F(y_t, \mathcal{E}_t),
\end{equation}
which penalizes claims that are not supported by the evidence. This term explicitly constrains generated text to remain consistent with the underlying model state. As a result, the report is encouraged to reflect actual system behavior rather than unsupported inferences.

\subsubsection{Confidence-Aware Conditioning}
The stage confidence score $c_t = \max_k p_{t,k}$ is incorporated into the prompt:
\begin{equation}
\mathcal{P}_t \leftarrow \mathcal{P}_t \cup \{c_t\}.
\end{equation}
This allows the language model to calibrate its phrasing according to uncertainty. For example, high-confidence situations may justify assertive wording, whereas low-confidence situations should lead to cautious descriptions. This helps align generated language with the model’s epistemic uncertainty.

\subsubsection{Length Regularization}
To ensure concise and actionable reports, DeepFaith employs a compactness loss, as defined in Equation~\eqref{L_compact}, to penalize outputs whose length exceeds a predefined budget $L_{\max}$, where $|y_t|$ denotes the length of the generated report. This constraint is particularly important in SOC environments, where analysts typically prefer short, information-dense summaries. By limiting unnecessary verbosity, the model is encouraged to produce focused reports that highlight only the most relevant information.

\subsection{Verification and Regeneration}
At inference time, the incident report is generated by maximizing the conditional likelihood under the language model:
\begin{equation}
y_t = \arg\max_y P_\theta(y \mid \mathcal{P}_t),
\end{equation}
where $\mathcal{P}_t$ is the evidence-grounded prompt and $P_\theta$ denotes the parameterized language model. In practice, this maximization can be approximated using decoding strategies such as beam search or nucleus sampling \cite{habibzadeh2025socsurvey}.

To further improve reliability, DeepFaith incorporates a post-generation verification step that evaluates the faithfulness score $F(y_t, \mathcal{E}_t)$, which quantifies the extent to which the generated report is supported by the structured evidence $\mathcal{E}_t$. If the faithfulness score falls below a predefined threshold $\tau_F \in [0,1]$, i.e., $F(y_t, \mathcal{E}_t) < \tau_F$, the generated report is flagged for regeneration, as indicated by the red dashed feedback loop in Fig.~\ref{fig:DeepFaith_framework}.  The threshold $\tau_F$ controls the strictness of the verification process and can be tuned according to application requirements. This mechanism acts as a safeguard against unsupported or hallucinated content, ensuring that only reliable and evidence-grounded reports are presented to analysts, which is critical in safety-sensitive SOC environments.

\subsection{Human-AI Collaboration in SOC}
DeepFaith reports support SOC workflows such as triage, mitigation, and investigation. By providing structured summaries, timelines, and rationales, the system improves situational awareness and reduces cognitive load. Confidence-aware expressions further help analysts calibrate trust, enabling effective human–AI collaboration.

\section{Case Study: Realistic APT Lateral Movement}
\subsection{Attack Scenario Overview}
We consider a stage space with $K=7$ discrete states, indexed as $k \in \{0,1,2,3,4,5,6\}$, where $k=0$ denotes the \textit{no-attack} (benign) state, and $k=1$ to $6$ correspond to six APT stages \cite{mitre_caldera}: reconnaissance, initial compromise, privilege escalation, lateral movement, command-and-control, and data exfiltration, respectively.

In this scenario, an attacker initially compromises a user workstation in the LAN zone (Host A) through credential leakage and attempts to move laterally to a database server in the server zone (Host B). The attack leverages credential dumping and remote service execution over SMB, while benign background activities are simultaneously present.

\subsection{Step-by-Step Attack Execution}
\textbf{Step 1: Initial Compromise and Credential Access.} 
The attacker executes a credential dumping script on Host A:
\begin{itemize}
\item {\footnotesize \texttt{/usr/bin/python3 cred\_dump.py}}
\end{itemize}
which accesses sensitive files such as:
\begin{itemize}
\item {\footnotesize\texttt{/etc/shadow}}
\end{itemize}

\textbf{Step 2: Credential Reuse and Remote Execution.} 
Using the extracted credentials, the attacker initiates:
\begin{itemize}
\item {\footnotesize\texttt{psexec.py -target HostB -u admin -p *****}}
\end{itemize}
This results in SMB authentication and remote execution from Host A (LAN) to Host B (server zone).

\textbf{Step 3: Provenance Graph Construction.} 
The telemetry is fused into a provenance graph $G_t$, where nodes represent processes ({\footnotesize\texttt{python3}}, {\footnotesize\texttt{psexec}}), files ({\footnotesize\texttt{/etc/shadow}}), and network sockets, and edges encode causal interactions across hosts and network zones.

\textbf{Step 4: Stage Estimation and Defense Action.} 
The GNN encodes $G_t$ into embedding $g_t$, and the stage estimator produces the following probability distribution over $K=7$ stages:
$p_t = [0.03,\, 0.04,\, 0.05,\, 0.06,\, 0.78,\, 0.03,\, 0.01]$.
The confidence score is $c_t = \max_k p_t^{(k)} = 0.78$, indicating that the system is highly confident the attack is in the lateral movement stage ($k=4$). Based on the inferred belief state and the stage estimation, the DRL-based policy $\pi_\theta$ selects a defense action $a_t$, which in this case corresponds to \textit{network interface isolation} ($a_{19}$ in the \textit{DeepStage} \cite{deepstage}). This action aims to contain the compromised host and prevent further lateral propagation within the enterprise network.

\textbf{Step 5: Explanation Generation.} 
The XAI evidence module extracts:
$G_t^*$: a critical subgraph capturing  
{\footnotesize\texttt{python3} $\rightarrow$ \texttt{/etc/shadow} $\rightarrow$ \texttt{psexec} $\rightarrow$ \texttt{SMB connection}},
$I_t$: key timestamps (e.g., {\footnotesize\texttt{10:32:15}} and {\footnotesize\texttt{10:33:02}}),
$A_t$: rationale recommending {\footnotesize\texttt{network interface isolation}}.

\textbf{Step 6: Evidence-Grounded Report Generation.} 
DeepFaith constructs prompt $\mathcal{P}_t$ from $\mathcal{E}_t$ and generates:

{\footnotesize
\begin{verbatim}
[DeepFaith SOC Report]
----------------------------------------
Stage: Lateral Movement (k=4)
Confidence: 0.78
[Timeline]
  10:32:15-Credential file access detected.
  10:33:02-Remote execution to internal server.
[Analysis]
  Credential reuse and unauthorized lateral 
  movement detected across network zones.
[Recommended Action]
  Isolate the affected host to prevent further spread.
----------------------------------------
\end{verbatim}
}

\textbf{Step 7: Faithfulness Verification.} 
The report is decomposed into claims $\phi(y_t)$ and matched against evidence $\psi(\mathcal{E}_t)$. Suppose: $F(y_t, \mathcal{E}_t) = 0.83$. Since $F(y_t, \mathcal{E}_t) \geq \tau_F = 0.8$, the report is accepted.

\textbf{Step 8: SOC Response.} 
The analyst isolates Host A, blocks SMB connections, resets credentials, initiates threat hunting.

Note that this behavior corresponds to the Lateral Movement tactic (TA0008) and Remote Services technique (T1021) in the MITRE ATT\&CK framework \cite{mitre_caldera}.

\section{Evaluation Setup}
\subsection{Enterprise Testbed and Data Sources}
Following our prior \textit{DeepStage}~\cite{deepstage}, experiments are conducted in a realistic enterprise testbed segmented into four zones: LAN, DMZ, server, and management. This setup reflects typical enterprise architectures and enables evaluation of multi-stage APT behavior across domains. Host telemetry is collected from endpoint monitoring tools (e.g., Auditd), while network events are captured by Zeek sensors. These data sources are normalized and fused into provenance graphs $G_t$, where nodes represent system entities (e.g., processes, files, sockets) and edges encode causal interactions such as execution and access.

To emulate realistic APT campaigns, we use CALDERA-driven adversarial playbooks~\cite{mitre_caldera}, covering all attack stages from reconnaissance to data exfiltration. Benign background activity is executed concurrently to introduce noise and reflect real SOC environments. The same attack-stage definitions, telemetry intervals, and graph construction settings as our prior \textit{DeepStage} are used for consistency. Results are averaged over 10 runs to account for stochastic variability.

\subsection{APT Defense and XAI Setup}
We follow the pipeline of our prior \textit{DeepStage}~\cite{deepstage} and \textit{DeepXplain}~\cite{deepxplain} frameworks. The provenance graph $G_t$ is encoded by a GNN into embedding $g_t$. A recurrent estimator produces stage distributions $p_t$, while a DRL agent generates belief states $b_t$ and defense actions $a_t$. The prior \textit{DeepXplain} module produces structured explanations $(G_t^*, I_t, A_t)$, which are integrated into the unified evidence representation $\mathcal{E}_t$ used by DeepFaith.

\subsection{Language Model and Implementation}
DeepFaith uses a large language model with structured prompting. We adopt LLaMA-2-13B~\cite{touvron2023llama2} as the backbone for evidence-grounded report generation. The model is conditioned on serialized evidence to produce faithful and interpretable outputs (see Table~\ref{tab:LLMConfiguration}). The serialization module converts $G_t^*$ into entity-relation triples and $I_t$ into ordered timestamps, forming structured textual inputs. Confidence scores $c_t$ are included in prompts to regulate linguistic uncertainty. A verification module computes $F(y_t, \mathcal{E}_t)$ and triggers regeneration if it falls below a threshold, ensuring outputs remain grounded in evidence. 

Note that in the current implementation, Eq.~(\ref{OptimizationProblem}) serves as a design objective, while its components are approximated through structured prompting, confidence-aware conditioning, compact decoding, and post-generation verification rather than end-to-end fine-tuning.

\begin{table}
\centering
\caption{DeepFaith Model and Module Configuration}
\large
\label{tab:LLMConfiguration}
\renewcommand{\arraystretch}{1.0}
\resizebox{\columnwidth}{!}{
\begin{tabular}{p{2.0cm} p{3.6cm} p{5.2cm}}
\hline
\textbf{Module} & \textbf{Component} & \textbf{Setting} \\
\hline
\multirow{3}{*}{Verification}
& Faithfulness metric & $F(y_t, \mathcal{E}_t)$ \\
& Threshold & $\tau_F = 0.8$ \\
Module & Regeneration & Triggered if $F < \tau_F$ \\
\hline
\multirow{3}{*}{Language}
& Model & LLaMA-2-13B~\cite{touvron2023llama2} \\
& Decoding & Beam search ($B = 5$) \\
Model & Maximum length & $L_{\max} = 256$ tokens \\
\hline
\multirow{3}{*}{Prompt}
& Prompt type & Structured prompt $\mathcal{P}_t$ \\
& Input fields & $\{s_t, \tau_t, \gamma_t, \alpha_t\}$ \\
Construction & Confidence input & $c_t$ appended to prompt \\
\hline
\multirow{3}{*}{Evidence}
 & Graph representation & Entity-relation triples from $G_t^*$ \\
& Temporal encoding & Ordered timestamps from $I_t$ \\
Serialization & Action explanation & Textual rationale from $A_t$ \\
\hline
\end{tabular}
}
\end{table}

\subsection{Baselines}
We compare the DeepFaith framework against three representative baselines that reflect common approaches for incident report generation in SOC environments. All baselines use the same underlying LLM to ensure fair comparison, differing only in input representation and prompting strategy.

\subsubsection{Template-Based Reporting} This baseline implements a rule-based reporting system using predefined templates. Specifically, structured inputs from $\mathcal{E}_t$ are directly mapped into fixed textual fields. For example, the stage is selected as $\hat{k}_t = \arg\max_k p_t^{(k)}$, timestamps are taken from $I_t$, and the selected action $a_t$ is converted into a predefined description. Similar template-based reporting is widely used in SIEM systems~\cite{behl2019siem}. 

\subsubsection{Vanilla LLM} This baseline uses the same LLM backbone (LLaMA-2-13B) but without structured evidence grounding. The model is prompted using raw or minimally processed inputs, i.e., our prior \textit{DeepXplain} outputs $(G_t^*, I_t, A_t)$. No structured serialization or grounding constraints are applied. This baseline follows standard LLM usage for text generation~\cite{brown2020gpt3}. 

\subsubsection{Chain-of-Thought (CoT) Prompting} This baseline extends the Vanilla LLM by incorporating reasoning-oriented prompts. The model first produces a reasoning chain (e.g., identifying suspicious processes, inferring attack stage), followed by a final report. This baseline follows the chain-of-thought prompting paradigm~\cite{wei2022cot}.

\section{Results and Analysis}

\subsection{Quantitative Comparison}
\begin{figure}[t]
\centering
\includegraphics[width=0.5\textwidth]{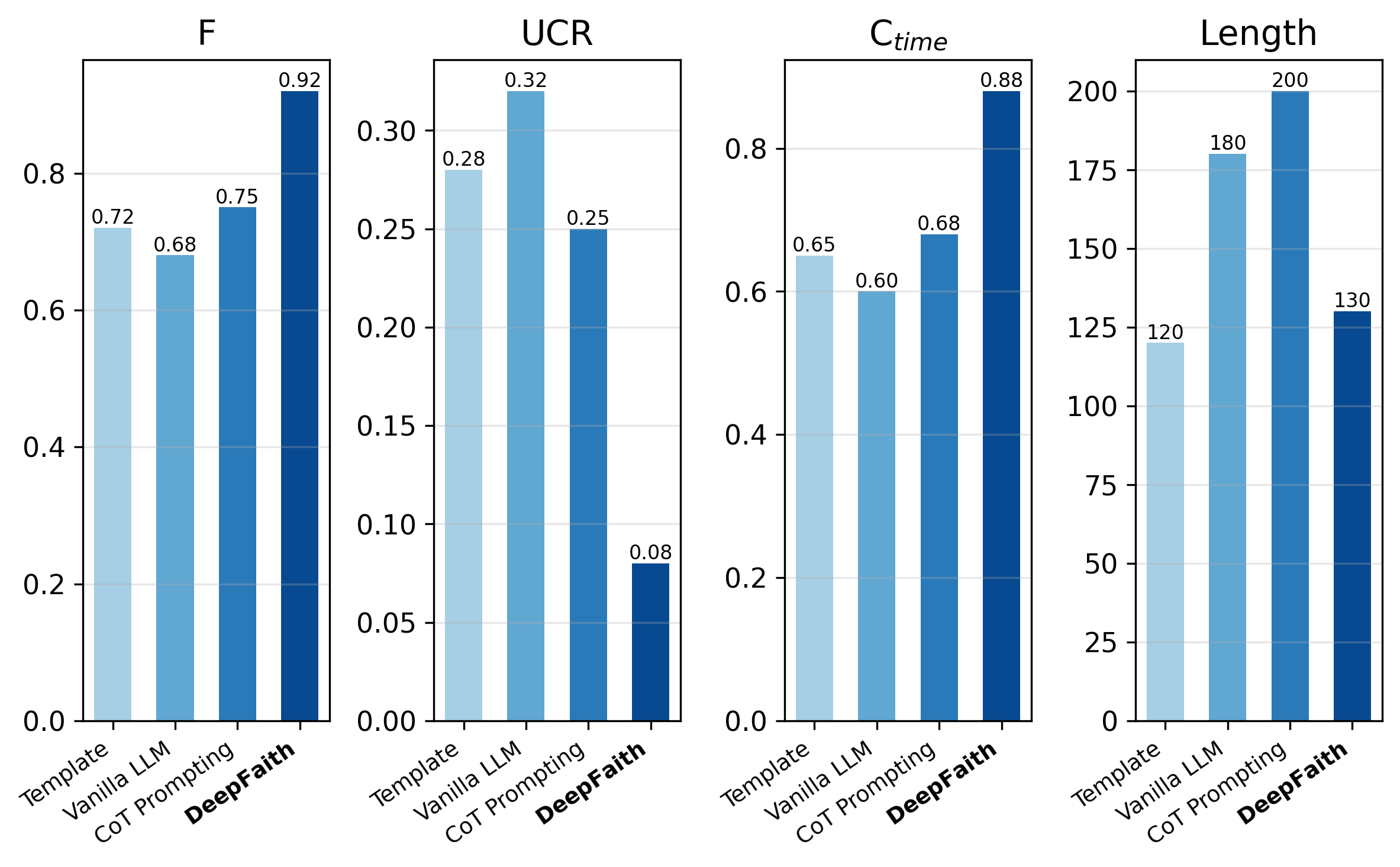}
\caption{Performance comparison across four methods.}
\label{fig:Performance_Comparison}
\end{figure}

Fig.~\ref{fig:Performance_Comparison} summarizes the performance of DeepFaith compared to three representative baselines across key evaluation metrics. We evaluate report quality using four complementary metrics: (\textit{i}) \textit{Faithfulness} ($F$), defined as the proportion of claims in the generated report that are supported by structured evidence $\mathcal{E}_t$; (\textit{ii}) \textit{Unsupported Claim Rate} (UCR), measuring the fraction of claims not grounded in $\mathcal{E}_t$; (\textit{iii}) \textit{Temporal Consistency} ($C_{\mathrm{time}}$), capturing alignment between generated timelines and important timestamps $I_t$; and (\textit{iv}) \textit{Report Length}, indicating conciseness. Together, these metrics evaluate correctness, completeness, coherence, and operational usability.

DeepFaith achieves the highest faithfulness ($F=0.92$), significantly outperforming Template-Based Reporting ($0.72$), Vanilla LLM ($0.68$), and CoT Prompting ($0.75$). This demonstrates that conditioning generation on structured evidence $\mathcal{E}_t$, combined with faithfulness-aware constraints, effectively ensures that generated reports remain aligned with model-derived information. In contrast, Vanilla LLM produces fluent but weakly grounded outputs due to reliance on implicit knowledge, while CoT Prompting improves reasoning coherence but lacks mechanisms to enforce consistency with structured evidence. These results highlight that reasoning alone is insufficient without explicit grounding.

This improvement is further reflected in the Unsupported Claim Rate (UCR), where DeepFaith achieves the lowest value ($0.08$), compared to Template ($0.28$), Vanilla LLM ($0.32$), and CoT ($0.25$). Since UCR directly measures hallucinated or unsupported content, this reduction confirms the effectiveness of combining structured prompting with post-generation verification. The verification mechanism acts as a safeguard that filters or regenerates outputs that deviate from $\mathcal{E}_t$, which is critical in safety-sensitive SOC environments.

DeepFaith also achieves the highest temporal consistency ($C_{\mathrm{time}} = 0.88$), indicating strong alignment between generated timelines and evidence-supported timestamps $I_t$. In contrast, Vanilla LLM ($0.60$) often produces incomplete or misordered event sequences, while CoT Prompting ($0.68$) provides only moderate improvements without explicit temporal grounding. These results show that incorporating temporal evidence directly into the generation process is essential for reconstructing coherent multi-stage attack progression.

In terms of conciseness, DeepFaith produces compact reports (130 tokens), comparable to Template-Based Reporting (120 tokens) and substantially shorter than Vanilla LLM (180 tokens) and CoT Prompting (200 tokens). This demonstrates that the compactness constraint effectively balances informativeness and brevity. Unlike template-based methods, which limit expressiveness, DeepFaith maintains rich contextual information while avoiding redundancy, resulting in more information-dense outputs.

\begin{figure}[t]
\centering
\includegraphics[width=0.5\textwidth]{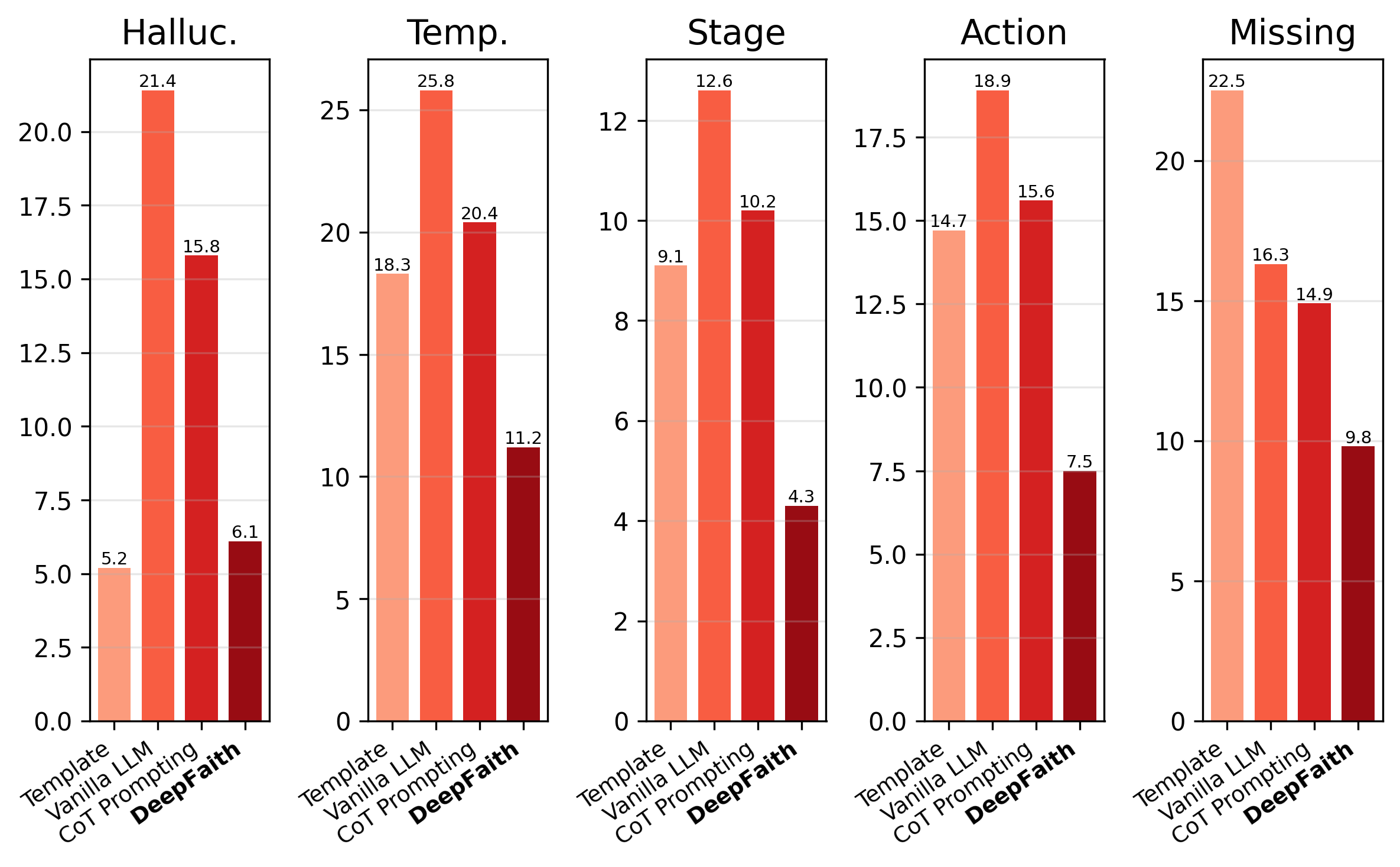}
\caption{Error type distribution across methods (\%).}
\label{fig:Error_Analysis}
\end{figure}

\subsection{Error Type Analysis}
To analyze limitations of generated reports, we perform error analysis across methods. For each report $y_t$, we extract claims $\phi(y_t)$ and compare them with evidence $\psi(\mathcal{E}_t)$ to identify errors affecting faithfulness, temporal consistency, and decision reliability. We define the following error categories: (\textit{i}) \textit{Hallucinated Event}: A claim $c \in \phi(y_t)$ that cannot be matched to any fact in $\psi(\mathcal{E}_t)$, indicating unsupported or fabricated content; (\textit{ii}) \textit{Temporal Error}: An event whose timestamp is inconsistent with $I_t$, including misordered events or missing temporal anchors; (\textit{iii}) \textit{Stage Misclassification}: The reported attack stage does not correspond to the maximum-probability stage in $p_t$; (\textit{iv}) \textit{Incorrect Action Rationale}: The generated explanation of action $a_t$ does not match the rationale encoded in $A_t$; and (\textit{v}) \textit{Missing Critical Evidence}: Key facts from $\psi(\mathcal{E}_t)$ (e.g., important entities or events in $G_t^*$) are not reflected in $\phi(y_t)$. We analyze 200 reports from multi-stage APT scenarios, compute error frequencies, and normalize them to obtain error rates. Fig.~\ref{fig:Error_Analysis} summarizes the results.

DeepFaith consistently reduces all error types compared to baselines. Hallucinated events drop from 21.4\% (Vanilla LLM) and 15.8\% (CoT) to 6.1\%, confirming that evidence-grounded prompting and verification suppress unsupported claims. Temporal errors are also lower (11.2\% vs.\ 25.8\%), showing that incorporating temporal evidence $I_t$ improves alignment. DeepFaith further achieves the lowest rate of incorrect action rationales (7.5\%), indicating that structured explanations $A_t$ guide decision-consistent reporting. Template-based reporting has low hallucination (5.2\%) but high missing evidence (22.5\%), reflecting limited expressiveness. Vanilla LLM produces more complete but less reliable outputs with high hallucination and temporal inconsistency, while CoT reduces some reasoning errors but still lacks explicit grounding, leading to unsupported or incomplete outputs.

Overall, this analysis shows that DeepFaith not only improves aggregate metrics such as faithfulness and temporal consistency, but also systematically reduces specific error types. These improvements directly contribute to more reliable and actionable reports for effective SOC decision-making.

\begin{figure}[t]
\centering
\includegraphics[width=0.5\textwidth]{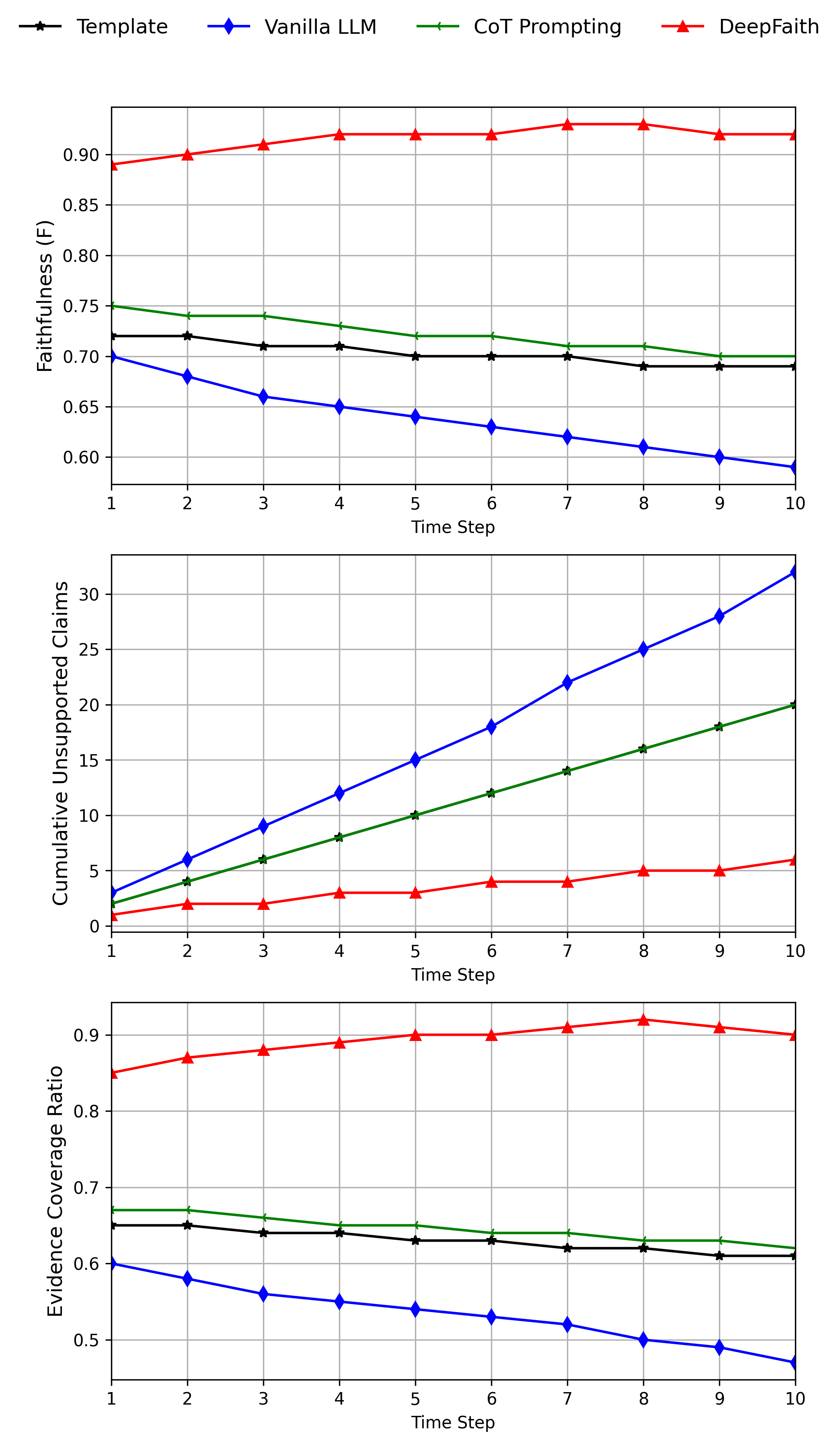}
\caption{Temporal analysis across multi-stage APT progression.}
\label{fig:Tempral_Metrics}
\end{figure} 

\subsection{Visualization of Temporal Metrics}
To complement Fig.~\ref{fig:Performance_Comparison}, Fig.~\ref{fig:Tempral_Metrics} shows temporal trends of three metrics: \textit{Faithfulness} ($F$), the proportion of claims supported by $\mathcal{E}_t$; \textit{Cumulative Unsupported Claims} (CUC), the accumulated number of unsupported claims over time; and \textit{Evidence Coverage Ratio} (ECR), the fraction of evidence in $\psi(\mathcal{E}_t)$ reflected in the generated report. 

DeepFaith maintains consistently high $F$, indicating stable grounding as attack complexity increases, while Vanilla LLM degrades and CoT shows moderate improvement. DeepFaith also slows the growth of CUC, demonstrating effective control of long-horizon error propagation. In addition, it achieves the highest ECR, capturing more complete structural, temporal, and action-level evidence. In contrast, Template-based reporting omits contextual details, while Vanilla LLM and CoT exhibit inconsistent evidence usage. Overall, DeepFaith provides more reliable reporting across all attack stages.

\begin{table}[t]
\centering
\caption{Ablation Study of DeepFaith}
\label{tab:ablation}
\small
\begin{tabular}{lcccc}
\hline
Method & $F \uparrow$ & UCR $\downarrow$ & $C_{\mathrm{time}} \uparrow$ & Length \\
\hline
w/o Serialization & 0.82 & 0.18 & 0.78 & 145 \\
w/o Prompt & 0.84 & 0.16 & 0.80 & 142 \\
w/o Faithfulness & 0.79 & 0.21 & 0.81 & 138 \\
w/o Confidence & 0.90 & 0.10 & 0.87 & 131 \\
w/o Temporal & 0.88 & 0.12 & 0.74 & 134 \\
w/o Verification & 0.86 & 0.14 & 0.84 & 136 \\
\textbf{DeepFaith} & \textbf{0.92} & \textbf{0.08} & \textbf{0.88} & \textbf{130} \\
\hline
\end{tabular}
\end{table}

\subsection{Ablation Study}
We evaluate six ablated variants of DeepFaith by removing one component at a time while keeping the same backbone, scenarios, and decoding setup: (\textit{i}) \textit{w/o Serialization}, (\textit{ii}) \textit{w/o Prompt}, (\textit{iii}) \textit{w/o Faithfulness}, (\textit{iv}) \textit{w/o Confidence}, (\textit{v}) \textit{w/o Temporal}, and (\textit{vi}) \textit{w/o Verification}. The full model includes all components. Table~\ref{tab:ablation} shows that all components contribute to performance. Removing serialization or structured prompting reduces faithfulness and increases unsupported claims, highlighting the importance of normalized evidence inputs. Removing faithfulness-aware conditioning leads to the largest drop in $F$ and highest UCR, confirming its key role in suppressing hallucination. Temporal removal causes the largest decline in $C_{\mathrm{time}}$, indicating the importance of temporal constraints. Disabling verification degrades all grounding-related metrics, showing that post-generation validation is critical. Confidence-aware conditioning has a smaller but consistent effect on overall quality. Overall, the full DeepFaith model achieves the best trade-off across all metrics.

\begin{table}[t]
\centering
\caption{Illustrative Evaluation of DeepFaith with Compact Open-Source LLM Backbones}
\label{tab:small_llm_eval}
\small
\begin{tabular}{lcccc}
\hline
Backbone & $F \uparrow$ & UCR $\downarrow$ & $C_{\mathrm{time}} \uparrow$ & Length\\
\hline
Llama-3.2-3B & 0.84 & 0.16 & 0.80 & 122 \\
Qwen2.5-3B & 0.88 & 0.12 & 0.84 & 126 \\
Gemma-3-4B-IT & 0.89 & 0.11 & 0.85 & 124 \\
Falcon3-7B & 0.90 & 0.10 & 0.86 & 127 \\
LLaMA-2-13B & 0.92 & 0.08 & 0.88 & 130 \\
\hline
\end{tabular}
\end{table}

\subsection{Evaluation with Compact Open-Source LLM Backbones}
To evaluate portability, we replace the default backbone with four compact open-source models: Qwen2.5-3B-Instruct \cite{yang2024qwen25}, Gemma-3-4B-IT \cite{gemma2024}, Llama-3.2-3B-Instruct \cite{touvron2024llama3}, and Falcon3-7B-Instruct \cite{falcon2023}. All models use the same DeepFaith pipeline, evidence serialization, prompt template, verification threshold, and decoding configuration with identical $L_{\max}$. This setting isolates the effect of backbone capacity while keeping the grounding mechanism unchanged.

Table~\ref{tab:small_llm_eval} shows a clear scaling trend across model sizes. The LLaMA-2-13B backbone achieves the best performance ($F=0.92$, UCR$=0.08$, $C_{\mathrm{time}}=0.88$), while compact models exhibit gradual degradation as model size decreases. Among smaller backbones, Falcon3-7B performs closest to the 13B model ($F=0.90$, UCR$=0.10$, $C_{\mathrm{time}}=0.86$), followed by Gemma-3-4B-IT and Qwen2.5-3B-Instruct with moderate drops in faithfulness ($0.89$ and $0.88$) and slightly higher UCR. Llama-3.2-3B-Instruct shows the lowest performance ($F=0.84$, UCR$=0.16$, $C_{\mathrm{time}}=0.80$), indicating reduced capability in maintaining grounding and temporal alignment. Nevertheless, all compact backbones maintain acceptable faithfulness and temporal consistency, suggesting that structured evidence conditioning reduces dependence on model scale.

Overall, these results show that while larger models provide incremental gains, DeepFaith achieves reliable performance even with lightweight backbones. This makes the framework suitable for deployment in resource-constrained SOC environments, where local inference, privacy preservation, and reduced computational cost are often important requirements.

\section{Conclusion}
This paper presented DeepFaith, an evidence-grounded framework for generating faithful and interpretable incident reports in autonomous APT defense systems. DeepFaith integrates structured evidence, constrained generation, and verification mechanisms to align reports with model-derived information. Experimental results show that DeepFaith consistently achieves higher faithfulness, lower unsupported claim rate, and better temporal consistency than template-based and LLM-based baselines, while producing concise reports. Error analysis further confirms significant reductions in hallucination and temporal inconsistencies. Ablation results demonstrate that grounding, temporal constraints, and verification are key to performance. DeepFaith also maintains strong performance with compact open-source LLMs, indicating that its effectiveness largely stems from structured evidence integration rather than model scale. Overall, DeepFaith enables more reliable and efficient SOC reporting. 
Future work includes extending DeepFaith toward interactive human-in-the-loop SOC systems, enabling bidirectional communication between analysts and autonomous defense models.

\section*{Acknowledgment}
This work has been performed in the framework of the SUSTAINET-Advance project, funded by the German BMFTR (ID:16KIS2280).

%\clearpage
%\balance
\bibliographystyle{ieeetr}
\bibliography{References.bib}

@inproceedings{deepstage,
  author    = {Trung V. Phan and others},
  title     = {DeepStage: Learning Autonomous Defense Policies Against Multi-Stage APT Campaigns},
  booktitle = {IEEE International Conference on Cyber Security and Resilience (CSR)},
  year      = {2026},
  note      = {[Accepted for Publication]. Available on https://arxiv.org/abs/2603.16969}
}

@inproceedings{deepxplain,
      title={DeepXplain: XAI-Guided Autonomous Defense Against Multi-Stage APT Campaigns}, 
      author    = {Trung V. Phan and others},
      booktitle = {IEEE Global Communications Conference (GLOBECOM)},
      year={2026},
      note      = {[Under Review]. Available on https://arxiv.org/abs/2603.21296}
}

@inproceedings{StageFinder,
  author    = {Trung V. Phan and others},
  title     = {Learning the APT Kill Chain: Temporal Reasoning over Provenance Data for Attack Stage Estimation},
  booktitle = {IEEE International Conference on Communications (ICC)},
  year      = {2026}
}

@ARTICLE{DeepAir,
  author={Trung V. Phan and others},
  journal={IEEE Transactions on Network and Service Management}, 
  title={DeepAir: Deep Reinforcement Learning for Adaptive Intrusion Response in Software-Defined Networks}, 
  year={2022},
  volume={19},
  number={3},
  pages={2207-2218},
  doi={10.1109/TNSM.2022.3158468}}

@misc{mitre_caldera,
  author       = {The MITRE Corporation},
  title        = {{MITRE Caldera: Automated Adversary Emulation Platform}},
  year         = {2024},
  url          = {https://github.com/mitre/caldera},
  version      = {5.0.0},
  note         = {Open-source adversary-emulation system used for breach-and-attack simulation of APT playbooks},
}

@misc{touvron2023llama2,
      title={Llama 2: Open Foundation and Fine-Tuned Chat Models}, 
      author={Hugo Touvron and others},
      year={2023},
      eprint={2307.09288},
      archivePrefix={arXiv},
      primaryClass={cs.CL},
      url={https://arxiv.org/abs/2307.09288}, 
      note={Available on https://arxiv.org/abs/2307.09288}
}

@article{alshamrani2019survey,
  title={A Survey on Advanced Persistent Threats: Techniques, Solutions, Challenges, and Research Opportunities},
  author={Alshamrani, Adel and others},
  journal={IEEE Communications Surveys \& Tutorials},
  volume={21},
  number={2},
  pages={1851--1877},
  year={2019}
}

@article{zhang2024survey,
  title={A Survey on Advanced Persistent Threat Detection: A Unified Framework, Challenges, and Countermeasures},
  author={Zhang, Bo and others},
  journal={ACM Computing Surveys},
  volume={57},
  number={},
  year={2024}
}

@article{nguyen2021drlcyber,
  title={Deep Reinforcement Learning for Cyber Security},
  author={Nguyen, Thanh Thi and others},
  journal={IEEE Transactions on Neural Networks and Learning Systems},
  volume={34},
  number={8},
  pages={3779--3795},
  year={2023}
}

@inproceedings{rehman2024flash,
  title={FLASH: A Comprehensive Approach to Intrusion Detection via Provenance Graph Representation Learning},
  author={Rehman, Mati Ur and others},
  booktitle={IEEE Symposium on Security and Privacy},
  year={2024}
}

@inproceedings{bilot2025simpler,
  title={Sometimes Simpler is Better: A Comprehensive Analysis of State-of-the-Art Provenance-Based Intrusion Detection Systems},
  author={Bilot, Tristan and others},
  booktitle={USENIX Security Symposium},
  year={2025}
}

@article{milani2024xrl,
  title={Explainable Reinforcement Learning: A Survey and Comparative Review},
  author={Milani, Stephanie and others},
  journal={ACM Computing Surveys},
  volume={56},
  number={7},
  year={2024}
}

@inproceedings{ying2019gnnexplainer,
  title={GNNExplainer: Generating Explanations for Graph Neural Networks},
  author={Ying, Rex and others},
  booktitle={Advances in Neural Information Processing Systems},
  year={2019}
}

@article{li2025gnnsurvey,
  title={Can Graph Neural Networks Be Adequately Explained? A Survey},
  author={Li, Xuan and others},
  journal={ACM Computing Surveys},
  year={2025}
}

@article{dai2024trustworthy,
  title={A Comprehensive Survey on Trustworthy Graph Neural Networks: Privacy, Robustness, Fairness, and Explainability},
  author={Dai, Enyan and others},
  journal={International Journal of Automation and Computing},
  year={2024}
}

@article{vyas2025realistic,
  title={Towards the Deployment of Realistic Autonomous Cyber Network Defence: A Systematic Review},
  author={Vyas, S. and others},
  journal={ACM Computing Surveys},
  year={2025}
}

@inproceedings{le2024automated,
  title={Automated APT Defense Using Reinforcement Learning and Attack Graph Risk-Based Situation Awareness},
  author={Le, Anh T. and others},
  booktitle={Workshop on Autonomous Cybersecurity},
  year={2024}
}

@inproceedings{thompson2024entity,
  title={Entity-Based Reinforcement Learning for Autonomous Cyber Defence},
  author={Thompson, S. and others},
  booktitle={Workshop on Autonomous Cybersecurity},
  pages={56--67},
  year={2024}
}

@misc{habibzadeh2025socsurvey,
      title={Large Language Models for Security Operations Centers: A Comprehensive Survey}, 
      author={Ali Habibzadeh and Farid Feyzi and Reza Ebrahimi Atani},
      year={2025},
      eprint={2509.10858},
      archivePrefix={arXiv},
      primaryClass={cs.CR},
      url={https://arxiv.org/abs/2509.10858},
      note={Available on https://arxiv.org/abs/2509.10858}
}

@article{jaffal2025llmcyber,
  title={Large Language Models in Cybersecurity: A Survey of Applications, Vulnerabilities, and Defenses},
  author={Jaffal, N. O. and others},
  journal={AI},
  volume={6},
  number={9},
  pages={216},
  year={2025}
}

@inproceedings{kramer2025incident,
  title={Integrating Large Language Models into Security Incident Response},
  author={Kramer, Diana and others},
  booktitle={Symposium on Usable Privacy and Security},
  year={2025}
}

@article{yang2025llmknowledge,
  title={A Comprehensive Survey on Integrating Large Language Models with Knowledge Bases},
  author={Yang, Wei and others},
  journal={Knowledge-Based Systems},
  year={2025},
  publisher={Elsevier},
  note={Survey on LLM and knowledge integration}
}

@inproceedings{structuredllm2026survey,
   title={Retrieval And Structuring Augmented Generation with Large Language Models},
   url={http://dx.doi.org/10.1145/3711896.3736557},
   DOI={10.1145/3711896.3736557},
   booktitle={Proceedings of the 31st ACM SIGKDD Conference on Knowledge Discovery and Data Mining V.2},
   publisher={ACM},
   author={Jiang, Pengcheng and others},
   year={2025},
   month=aug, pages={6032–6042},
   collection={KDD ’25} }

@article{behl2019siem,
  title={A Review of SIEM Systems},
  author={Behl, Abhishek and others},
  journal={Journal of Information Security},
  year={2019}
}

@article{brown2020gpt3,
  title={Language Models are Few-Shot Learners},
  author={Brown, Tom B. and others},
  journal={NeurIPS},
  year={2020}
}

@article{wei2022cot,
  title={Chain-of-Thought Prompting Elicits Reasoning in Large Language Models},
  author={Wei, Jason and others},
  journal={NeurIPS},
  year={2022}
}

@misc{yang2024qwen25,
      title={Qwen2.5 Technical Report}, 
      author={An Yang and others},
      year={2025},
      eprint={2412.15115},
      archivePrefix={arXiv},
      primaryClass={cs.CL},
      url={https://arxiv.org/abs/2412.15115},
      note={Available on https://arxiv.org/abs/2412.15115}
}

@misc{touvron2024llama3,
      title={The Llama 3 Herd of Models}, 
      author={Aaron Grattafiori and others},
      year={2024},
      eprint={2407.21783},
      archivePrefix={arXiv},
      primaryClass={cs.AI},
      url={https://arxiv.org/abs/2407.21783}, 
      note={Available on https://arxiv.org/abs/2407.21783}
}

@misc{gemma2024,
      title={Gemma: Open Models Based on Gemini Research and Technology}, 
      author={Gemma Team and others},
      year={2024},
      eprint={2403.08295},
      archivePrefix={arXiv},
      primaryClass={cs.CL},
      url={https://arxiv.org/abs/2403.08295},
      note={Available on https://arxiv.org/abs/2403.08295}
}

@misc{falcon2023,
      title={The Falcon Series of Open Language Models}, 
      author={Ebtesam Almazrouei and others},
      year={2023},
      eprint={2311.16867},
      archivePrefix={arXiv},
      primaryClass={cs.CL},
      url={https://arxiv.org/abs/2311.16867}, 
      note={Available on https://arxiv.org/abs/2311.16867}
}

\end{document}